%% file: paper.tex
\documentclass[conference]{IEEEtran}

\usepackage{amsmath}
\usepackage{amsfonts}
\usepackage{multirow}
\usepackage{cite}
\usepackage{color}
\usepackage{pslatex}
\usepackage{xspace}
\usepackage{amssymb}
\usepackage{amsbsy}
\usepackage{eucal}
\usepackage{mathrsfs}
\usepackage{multirow}
\usepackage{setspace}
\usepackage{bm}

\input{def}

\ifCLASSOPTIONcompsoc
  \usepackage{cite}
\else
  \usepackage{cite}
\fi

\ifCLASSINFOpdf
  \usepackage[pdftex]{graphicx}
  \DeclareGraphicsExtensions{.pdf,.jpeg,.png}
\else
  \usepackage[dvips]{graphicx}
   \DeclareGraphicsExtensions{.eps}
\fi

\ifCLASSOPTIONcompsoc
\else
\fi

\ifCLASSOPTIONcompsoc
\else
\fi

\begin{document}

\title{Toward Using Surrogates to Accelerate Solution of Stochastic Electricity Grid Operations Problems}
\author{\IEEEauthorblockN{Cosmin Safta, Richard L. Chen, Habib N. Najm,
               Ali Pinar}
\IEEEauthorblockA{Sandia National Laboratories,     \\
 Livermore, CA 94551 \\
 Email: \{csafta,rlchen,hnnajm,apinar\}@sandia.gov}
\and 
\IEEEauthorblockN{Jean-Paul~Watson} 
\IEEEauthorblockA{Sandia National Laboratories, \\
Albuquerque, NM 87185 \\
Email: jwatson@sandia.gov
}
}

\maketitle

\begin{abstract}
Stochastic unit commitment models typically handle uncertainties in forecast
demand by considering a finite number of realizations from a stochastic process
model for loads. Accurate evaluations of expectations or higher moments for the
quantities of interest require a prohibitively large number of model
evaluations. In this paper we propose an alternative approach based on using
surrogate models valid over the range of the forecast uncertainty. We
consider surrogate models based on Polynomial Chaos expansions, constructed
using sparse quadrature methods. Considering expected generation cost, we 
demonstrate the approach can lead to several orders of magnitude reduction 
in computational cost relative to using Monte Carlo sampling on the original model, 
for a given target error threshold.
\end{abstract}

\begin{IEEEkeywords}
Stochastic Unit Commitment, Monte Carlo Sampling, Polynomial Chaos Expansion
\end{IEEEkeywords}

\IEEEpeerreviewmaketitle

\section{Introduction}

\IEEEPARstart{U}{nit} commitment (UC) is the fundamental process of
scheduling thermal generating units in advance of operations in the
electric power grid \cite{carrionarroyo06}. The objective is to
minimize overall production costs to satisfy forecasted demand for
electricity, while respecting constraints on both transmission (e.g.,
thermal limits) and generator infrastructure (e.g., ramping
limits). Economic dispatch (ED) is a closely related 
operations problem, in which cost minimization is performed to
identify an optimal set of power output levels for a fixed set of
active thermal generating units. UC and ED are respectively formulated
as a mixed-integer and linear optimization problems, and solved using
commercial solvers. Despite improvements in forecasting technology,
next-day demand predictions are imperfect, with errors on average in
the 1-3\% range and exceeding 10\% on specific days \cite{is one}.
To account for such inaccuracies, reserve margins are universally
imposed in UC. These margins implicitly deal with uncertainty in load
forecasts, by ensuring there is sufficient generation capacity
available to meet unexpectedly high demand during operations. 

An alternative approach to dealing with forecast errors in UC is to
\emph{explicitly} model the load uncertainty, typically via a finite
set of sampled realizations from a stochastic process model of
load. This approach results in a stochastic UC model (SUC), in which
the objective typically is to minimize the expected cost across the
load scenarios \cite{ruizetal09,takritietal96}. By explicitly
representing the inherent uncertainty in load forecasts, a SUC
solution ensures sufficient flexibility to meet a range of potential
load realizations during next-day operations. Further, by explicitly
representing uncertainty, reliance on reserve margins is reduced,
yielding less costly solutions than those obtained for the
deterministic UC problems. While not considered here, we note that the
problems induced by increasing rates of renewables (e.g., wind and
solar) generation penetration accentuates the differences between
stochastic and deterministic UC problems, due to
increased errors in the next-day forecasts relative to load. While
conceptually appealing, the computational difficulty of stochastic UC
is well-known \cite{papavasiliou11}, such that it is not presently
used in practice. The difficulty is primarily driven by the number of
forecast samples required to achieve high-quality, robust solutions. 

Uncertainties such as those found in stochastic UC are ubiquitous in
both power systems operations and planning, and the importance of
credibly accounting for them is well-recognized. However, the lack of
advanced methods to handle uncertainty  and the limitations of
scenario-based approaches have led researchers  to seek
alternatives. For example, Thiam and DeMarco~\cite{Thiam:2010} argue:
``{\it Simply put, when uncertainty is credibly accounted for such
  methods yield solutions for economic benefit of a transmission
  expansion in which the ``error bars" are often larger than the
  nominal predicted benefit.}"  Instead, Thiam and
DeMarco~\cite{Thiam:2010} propose an ``oblivious approach" to
transmission expansion that does not take into account the
uncertainties in the inputs. We agree that an oblivious approach is
more credible than simply ignoring large error bars in
estimations. However, we posit that with proper modeling and  sampling
algorithms, the errors incurred in such analyses can be  drastically
reduced. Of course, it is not possible to change the nature of
uncertainties, such that if uncertainties are so large that they fail
to provide significant information, oblivious approaches may be
appropriate. However, as we argue in this paper, it is possible to
reduce additional uncertainties and inefficiencies introduced due to
poor modeling and sampling.

In this paper, we propose to adopt advanced modeling and sampling techniques
from the uncertainty quantification (UQ) community, and leverage them to impact
power systems operations problems such as stochastic UC and ED. Such techniques
have been successfully applied in many areas of computational science and
engineering, with great success~\cite{Najm:2009a}.  Most studies analyzing
uncertain power system operations problems generate forecast scenarios by
drawing random samples from a stochastic process using standard Monte Carlo (MC)
techniques. In this paper, we consider instead an alternative approach based on
using Polynomial Chaos~\cite{Ghanem:1991} expansions, built using sparse
quadrature methods, as surrogate models valid over the range of the forecast
uncertainty. We demonstrate that our approach yields a one to two order of
magnitude reduction in the number of samples required to estimate expected
generation cost, relative to MC, for a given target error threshold. Consequently, 
our approach has the potential to dramatically drop the computational difficulty of
stochastic UC and ED, significantly reducing the barriers to its use in
practice.

The remainder of this paper is organized as follows. We briefly
introduce our stochastic UC and ED formulations in Section~\ref{sec:suc}, to
provide context for our research. In Section~\ref{sec:estimation}, we
detail our surrogate models of load for stochastic UC / ED, based on 
Polynomial Chaos expansion. We then empirically analyze the accuracy
of our surrogates on standard IEEE test problems in 
Section~\ref{sec:results}, and conclude in Section~\ref{sec:conclusion}.

\section{Stochastic Commitment and Dispatch}
\label{sec:suc}

\IEEEPARstart{W}e now describe a generic formulation of the stochastic UC problem. 
Our formulation is based on the deterministic mixed-integer linear 
UC formulation introduced by \cite{carrionarroyo06}.  Let $G$ and $T$ 
denote the index sets of thermal generating units and time periods, 
respectively. We abstractly define the set of unit commitment constraints 
(i.e., operational and physical constraints on physical units) as $\mathcal
X$ and let $\boldsymbol x$ denote the vector of unit commitment decisions. 
The stochastic UC problem is then given as follows:

\begin{subequations}\label{2_stage}
\begin{align}
\min_{\boldsymbol {x} } \quad  &c^u(\boldsymbol x) +
c^d(\boldsymbol x) + \overline Q(\boldsymbol x) \\
\textmd{s.t.} \quad & \boldsymbol x \in \mathcal X,\\
& \boldsymbol x \in \{0,1\}^{|G|\times |T|}
\end{align}
\end{subequations}

The objective terms $c^u(\boldsymbol x)$ and $c^d(\boldsymbol x)$
represent generating unit start-up and shut-down costs, respectively, 
and $\overline Q(\boldsymbol x)$ denotes the expected generation
cost.  The UC constraints prescribed by $\mathcal X$ include minimum on-off requirements and
linearization of startup and shutdown costs.

We treat the loads $D^t$ for all $t \in T$ as random variables (RVs).
We begin by setting up a requisite theoretical framework as follows.
Define the probability space $(\Omega, \Sfr, P)$, where $\Omega$ is a sample
space, $\Sfr$ is a $\sigma$-algebra on $\Omega$, and $P$ is a probability
measure on $(\Omega,\Sfr)$. Further, defining the \emph{germ}
$\xivec=\{\xi_1,\xi_2,\ldots,\xi_{|T|}\}$ as a set of independent identically distributed
(\emph{iid}) RVs in $L_2(\Omega,\Sfr,P)$, to be further specified
below, we focus on the probability space $(\Omega,\Sfr_\xivec,P)$ employing the
sigma algebra generated by $\xivec$. We define the uncertain loads as RVs
$D^t(\omega):\Omega\rightarrow\mathbb{R}$ in
$L_2(\Omega,\Sfr_\xivec,P)$, such that we may write, by construction, 
$D^t:=D^t(\xivec(\omega))$, $\forall t\in T$.

Given the uncertain loads expressed as RVs, the corresponding generation cost 
$Q(\xvec,\xivec(\omega))$ is similarly uncertain/random. The
\emph{expected} generation cost, denoted $\overline Q(\boldsymbol x)$, is
defined as 
\begin{subequations}\label{buc}
\begin{align}
\overline Q(\boldsymbol x) = \mathrm{E}_{\boldsymbol \xi} Q(\boldsymbol x, \xivec(\omega))
\end{align}
\end{subequations}
\noindent and the uncertain (multi-period) \emph{economic dispatch problem} under a 
fixed unit commitment $\boldsymbol x$ is given by

\smallskip
\noindent $Q(\boldsymbol x, \xivec(\omega)) =$
\begin{subequations}\label{eq:sed}
\label{eq:rec_obj}
\begin{align}
\min_{\boldsymbol {p, q}} \quad & \sum_{t \in T} \sum_{g \in G} c_g^P(p_g^t) + \sum_{t \in T} Mq^t \label{ed_mod_obj}\\
\textmd{s.t.}\quad & \sum_{g \in   G}  p_g^t - q^t = D^t(\xi_t(\omega)), \quad \forall t \in T \label{ed_mod_bal}\\
&\underline P_g x_g^t \leq  p_g^t \leq \overline P_g x_g^t, \quad   \forall g \in G, t \in T \label{ed_mod_gen_lim} \\
&p_g^{t} -  p_g^{t-1} \leq RU(x_g^{t-1}, x_g^t), \quad  \forall g \in G, t \in T \label{ed_mod_ru} \\
&p_g^{t-1}-  p_g^t\leq RD(x_g^{t-1}, x_g^t), \quad  \forall g \in G, t \in T \label{ed_mod_rd}.
\end{align}
\end{subequations}

\noindent where 
\begin{align*}
&RU(x_g^{t-1}, x_g^t) = R_g^u x_g^{t-1} + {S}_g^u (x_g^t-x_g^{t-1}) + \overline P_g ( 1-x_g^t) \\
&RD(x_g^{t-1}, x_g^t) = R_g^d x_g^{t}+S_g^d (x_g^{t-1}-x_g^t)  +\overline P_g(1-x_g^{t-1}) 
\end{align*}
\noindent and $R_g^u/R_g^d$, $S_g^u/S_g^d$ represent nominal
ramp-up/ramp-down rates, startup/shutdown ramp rates, respectively. 

Note that because the loads are RVs, all solution variables are necessarily
RVs. For brevity of notation, we have only included the explicit dependence
on $\xivec(\omega)$ in $D^t$ and $Q$, thereby emphasizing the randomness of
inputs/output of interest only.

The optimization objective in stochastic ED is to minimize the expected total
production and loss-of-load costs. The first term in \eqref{ed_mod_obj}
represents total production cost.  The second term represents the
loss-of-load penalty, where $q^t$ is the unit (e.g., MW) of load unsatisfied
in period $t$.  Typically, the load-shedding penalty is equal to a 
large number $M$. Constraints \eqref{ed_mod_bal}-\eqref{ed_mod_rd}
specify operational constraints, and include (in order):
power balance at each period \eqref{ed_mod_bal}; lower and upper
bounds for committed generation unit output levels
\eqref{ed_mod_gen_lim}; and generation ramp-up and ramp-down
constraints for pairs of consecutive time periods \eqref{ed_mod_ru}
and \eqref{ed_mod_rd}.

A quadratic production cost function, given below, is typically employed in scheduling electricity
grid operations.

\begin{align}
c^P_g(p_g^t) = a_g x_g^t + b_g p_g^t + c_g (p_g^t )^2 \label{quad_prod_cost}
\end{align}

Equation \eqref{quad_prod_cost} can be accurately approximated by a set 
of piecewise linear segments.  For conciseness, we omit these standard
linearization steps.  For a detailed treatment on the linearization of the
quadratic cost function please refer to \cite{carrionarroyo06}.

The stochastic ED problem is embedded as a sub-problem in the stochastic
UC problem. The high-level context in stochastic UC is the presence of 
uncertainty in future loads (and, more generally, renewables and system 
component failures). In stochastic UC, the first-stage decisions are the 
unit commitment selections $\boldsymbol x$, and the objective is to 
minimize the expected generation costs. In the second (recourse) decision
stage, uncertain loads result in uncertain recourse decisions for the dispatch
variables $\boldsymbol p$ and $\boldsymbol q$, and associated generation 
and load shedding costs. First-stage unit commitment decisions
are determined by taking their future impacts into consideration.  These future 
impacts are quantified by the recourse function $\overline Q(\boldsymbol x)$,
which computes the expected value of generation cost for a given unit commitment 
$\boldsymbol x$. 

We can \emph{estimate} the expected generation cost by using a finite number of
load realizations (i.e., scenarios) $s \in \mathcal S$ sampled from the 
joint density $p(\boldsymbol D)$, where
$\boldsymbol D=\{D^1, D^2, \ldots, D^{|T|}\}$.
Defining $\rho \equiv 1/|\mathcal S|$, formulation \eqref{2_stage} 
can be rewritten as:

\begin{subequations}\label{eq:2_stage2}
\begin{align}
\min_{\boldsymbol {x}} \quad  &c^u(\boldsymbol x) +
c^d(\boldsymbol x) + \rho \sum_{s \in \mathcal S} Q(\boldsymbol x,s) \label{eq:qx}\\
\textmd{s.t.} \quad & \boldsymbol x \in \mathcal X,\\
& \boldsymbol x \in \{0,1\}^{|G|\times |T|}
\end{align}
\end{subequations}
where
\noindent $Q(\boldsymbol x,s) =$
\begin{subequations}
\label{eq:ed_mod2}
\begin{align}
\min_{\boldsymbol {p, q}} \quad & \sum_{t \in T} \sum_{g \in G} c_g^P(p_g^t) + \sum_{t \in T} Mq^t \label{ed_mod2_obj}\\
\textmd{s.t.}\quad & \sum_{g \in   G}  p_g^t - q^t = D^t_s, \quad \forall t \in T \label{ed_mod2_bal}\\
&\underline P_g x_g^t \leq  p_g^t \leq \overline P_g x_g^t, \quad   \forall g \in G, t \in T \label{ed_mod2_gen_lim} \\
&p_g^{t} -  p_g^{t-1} \leq RU(x_g^{t-1}, x_g^t), \quad  \forall g \in G, t \in T \label{ed_mod2_ru} \\
&p_g^{t-1}-  p_g^t\leq RD(x_g^{t-1}, x_g^t), \quad  \forall g \in G, t \in T \label{ed_mod2_rd}.
\end{align}
\end{subequations}

Formulation \eqref{eq:2_stage2} represents an \emph{extensive form} of the stochastic UC 
problem, based on $|\mathcal{S}|$ sampled scenarios of load realizations. Formulation 
\eqref{eq:sed} can be similarly discretized. 

\section{Accurate estimation with limited samples}
\label{sec:estimation}

The typical scenario sampling approach described above uses
Monte Carlo (MC) sampling to approximate an integration, thereby estimating an
expectation.  While  MC algorithms are commonly used for their convenience and
robustness, their poor convergence rate is well-known.  The MC estimate of the
expectation has error
\begin{equation}
\label{eq:error}
\mathrm{V}[Q(\boldsymbol x, \xivec)]/\sqrt{|S|},
\end{equation} 
where $\mathrm{V[Q]}$ denotes the variance of the RV $Q$.  Given the significant
additional complexity incurred by including stochasticity in the optimization
problem, a stochastic  formulation becomes advantageous relative to a deterministic
formulation when the variance is large.  Hence, accurate estimation of the
expectation is not only an academic exercise but is important in practice.

According to Eq.~(\ref{eq:error}), accurate estimation can be achieved by increasing the 
number of samples. However, a linear decrease in error requires a quadratic increase
in the number of samples, which can quickly render the stochastic optimization
problem intractable. This illustrates the limitation of MC
algorithms in providing accurate estimations; while they are convenient, they are
not efficient. 

Similar problems arise in uncertainty quantification (UQ) for computational
science in general, where each sample point typically corresponds to a full
simulation.  Here, estimation accuracy is of paramount
importance, as simulation results may lead to scientific discoveries  or
high-impact policy decisions. The need for accurate estimation of uncertain
model outputs, along with the prohibitive cost of MC samples, have lead to the
development of efficient alternatives to MC methods in UQ. In this paper, we
illustrate how we adopt such methods for stochastic optimization problems
in power systems.  The details of the proposed method will be explained 
subsequently. The key utility of the proposed approach is that it enables, in a
preprocessing step, \emph{efficient} construction of a compact, accurate, and
computationally inexpensive representation of the input-output map of the
uncertain system. This representation is then used as a \emph{surrogate} for
the dependence of select model-output quantities of interest (QoIs) on uncertain
model inputs. Using this surrogate, rather than the original system governing
equations, enables accurate estimation of uncertain QoIs, and associated
expectations, with minimal costs beyond those of surrogate construction.

\subsection{Representation of uncertainty using Polynomial Chaos} 

Given the formulation in Eq.~(\ref{eq:rec_obj}) with uncertain/random loads
leading to uncertain/random generation costs, we employ efficient UQ methods
that rely on \emph{functional representations} of random variables.  Specifically,
we use Polynomial Chaos (PC) expansions. A brief description of PC is presented
below. For an in-depth description, the reader is referred to a series of
publications on this topic~\cite{Wiener:1938,Ghanem:1991,Janson:1997,Xiu:2002c}.

Considering the above defined \emph{germ} $\xivec$ and the associated
probability space $(\Omega,\Sfr_\xivec,P)$, any RV $X: \Omega \rightarrow
\mathbb{R}$, where by construction $X\in L_2(\Omega,\Sfr_\xivec,P)$, can be
written as a PC expansion (PCE):
\be
X(\omega) = X(\xivec(\omega)) = \sum_{k=0}^{\infty} \alpha_k \Psi_k(\xivec)
\label{eq:pcedef}
\ee
where the basis functions $\Psi_k$ are multivariate
polynomials\footnote{Generally, other, non-polynomial basis functions can be
used, but we restrict ourselves here, without loss of generality, to the most
common polynomial-based usage.} that are orthogonal, by construction, with respect to the density of
$\xivec$. Thus
\be
\langle \Psi_i\Psi_j\rangle =
\int\Psi_i(\xi)\Psi_j(\xi)dP(\xi)=\delta_{ij}\langle \Psi_i^2\rangle
\ee
where $\delta_{ij}$ is Kronecker's delta. 
Further, given this orthogonality, we have
\be
\alpha_k = \frac{\langle X\Psi_k \rangle}{\langle\Psi_k^2\rangle}
\ee
where the inner product is defined, for any RV $Z(\xivec)$, by the Galerkin projection
\be
\langle Z \rangle = \int Z(\xivec) p_\xivec(\xivec) d\xivec .
\label{eq:proj}
\ee
Moreover, the $\Psi_k$ are products of univariate polynomials, namely
$\Psi_k(\xivec) =\psi_{k_1}(\xi_1)\cdots \psi_{k_n}(\xi_n)$, where $n=|T|$. In a practical
computational context, one truncates the PCE to order $p$.  The number of terms
in the resulting finite PCE
\be
X \approx \sum_{k=0}^{P} \alpha_k \Psi_k(\xivec)
\ee
is given by $P+1 = { (n+p)! }/{n!p!}$. We dispense with the $\approx$ symbol in the 
remainder of this paper, employing for any RV $X(\xivec)$ its truncated PCE
\be
X = \sum_{k=0}^{P} \alpha_k \Psi_k(\xivec).
\label{eq:pce}
\ee
Generalized PC (gPC) expansions have been developed by~\cite{Xiu:2002c} using a
broad class of orthogonal polynomials in the "Askey" scheme~\cite{Askey:1985}.
Each family of polynomials corresponds to a given choice of distribution for the
$\xi_i$ and is, by construction, orthogonal with respect to the density of $\xi_i$.
In general, the most useful choices for $(\xivec,\Psi)$ are uniform RVs with
Legendre polynomials and normal RVs with Hermite polynomials.

\subsection{Surrogate Construction}

We employ Legendre-Uniform (LU) PC, as it is most useful for purposes of
surrogate construction. Further, in this particular context, i.e.,
explicitly for surrogate construction, a key first step is to define the input
random variables as \emph{iid} uniform over their ranges of interest.  This does
\emph{not} restrict the utility of the approach to \emph{iid} uniform load
distributions. Rather, the uniform assumption is simply to ensure \emph{uniform}
\emph{accuracy} in the surrogate over the range of loads variability.  Once the
surrogate is available, providing effectively the input-output map, any
$p(\boldsymbol D)$ can be employed, as is further outlined below.

Since, by construction, for LU PC, $\xi_t\stackrel{iid}{\sim} U(-1,1),\ \forall t\in T$, defining
$D^t\stackrel{iid}{\sim} U(D^t_{\mathrm{min}},D^t_{\mathrm{max}})$, we have the PCE for $D^t$ given simply by
\be
D^t = \xi_t \frac{D^t_{\mathrm{max}}-D^t_{\mathrm{min}}}{2}
+ \frac{D^t_{\mathrm{max}}+D^t_{\mathrm{min}}}{2}, \quad \forall t\in T.
\ee
In this context we employ Eq.~(\ref{eq:pce}) to represent $Q(\bm{x},\xivec)$ with a truncated LU PCE
\begin{equation}
Q_\mathrm{PC}(\bm{x},\xivec) = \sum_{k=0}^P c_k(\bm{x})\Psi_k(\xivec),
\label{eq:qpce}
\end{equation}
where $\Psi_k(\xivec)$ are $n$-variate Legendre polynomials ($n=|T|$).
The coefficients $c_k$ depend on
the discrete variable $\bm{x}$, hence separate PCE approximations for $Q$
will be constructed for each instance of $\bm{x}$. Given Eq.~(\ref{eq:proj}), we have
\be
c_k(\bm{x}) = \frac{\langle Q\Psi_k\rangle}{\langle \Psi_k^2\rangle}=
\frac{1}{\langle \Psi_k^2\rangle}\int_{[-1,1]^n} Q(\bm{x},\xivec)\Psi_k(\xivec)d\xivec.
\label{eq:gproj}
\ee
where we have used $p_\xivec(\xivec)=1$, for $\xivec\in[-1,1]^n$. 

Given $Q_\mathrm{PC}(\boldsymbol x,\xivec)$,
then for the given $D^t \stackrel{iid}{\sim} U(D^t_\mathrm{min},
D^t_\mathrm{max})$, $\forall t\in T$, we have immediately that
\be
{\overline Q}(\bm{x})=\mathrm{E}_\xivec[Q(\bm{x},\xivec)]=\langle Q(\bm{x},\xivec) \rangle = c_0,
\ee
being the solution of the stochastic ED problem, as required for the stochastic UC problem specified in
Eq.~(\ref{2_stage}). 

Beyond this, however, the PCE $Q_\mathrm{PC}(\boldsymbol x, \xivec)$ can be used as a
\emph{surrogate} for the ED problem solution $Q(\bm{x},\xivec)$. Specifically,
$Q_\mathrm{PC}(\bm{x},\xivec(\bm{D}_s)) \approx Q(\boldsymbol x, s)$ (see Eq.~\ref{eq:ed_mod2})
for any demand $\boldsymbol D_s\in \mathbb{D}$ where $\mathbb{D}=\{\boldsymbol D |
D^t\in[D^t_{\mathrm{min}},D^t_{\mathrm{max}}] \, \forall t\in T\}$, employing 
\be
\xi_t = \frac{2 D^t - (D^t_{\mathrm{max}}+D^t_{\mathrm{min}})
}{D^t_{\mathrm{max}}-D^t_{\mathrm{min}} }, \quad \forall t\in T.
\label{eq:D2xi}
\ee
Thus, for any arbitrary $p(\boldsymbol D)$, the PCE $Q_\mathrm{PC}(\boldsymbol
x,\xivec(\boldsymbol D))$ can be used to \emph{efficiently} provide samples
$Q(\boldsymbol x,s)$ given random samples ${\boldsymbol D}_{s}\sim p(\boldsymbol
D)$, as long as ${\boldsymbol D}_s\in \mathbb{D}$.  Thus, the MC estimation of
the expectation $\overline Q = \rho^{-1}\sum_{s\in\mathcal S} Q(\boldsymbol x,
s)$ in Eq.~(\ref{eq:qx}) can be done with arbitrarily large $|\mathcal S|$ given
the low cost of samples. Alternatively, for any $p(\boldsymbol D)$, one can
employ the PCE for $\boldsymbol D(\etavec)$, where $\etavec$ is suitable PC
germ, and use sparse-quadrature methods outlined below to evaluate projection
integrals -- but now using as forward model the surrogate
$Q_\mathrm{PC}(\boldsymbol x, \xivec ( \boldsymbol D(\etavec)))$, to arrive at
the PCE for $Q(\boldsymbol x, \etavec)$, from which one easily has $\overline Q
\equiv c_0$, as long as the sparse-quadrature samples $\boldsymbol D(\etavec_q)
\in \mathbb{D}$.  The only remaining issue is the evaluation of the projection
integrals in Eq.~(\ref{eq:gproj}), which we discuss below.

\subsection{Evaluation of the Projection Integrals}

Several methods can be employed to evaluate the projection integrals in
Eq.~(\ref{eq:gproj}). MC methods can be used in principle, but are impractical
given their slow convergence rate. Alternatively, for smooth
integrands, and particularly in low-moderate dimensional problems, sparse
quadrature methods~\cite{Smolyak:1963,Gerstner:1998,Conrad:2013} can provide
highly accurate results with smaller numbers of deterministic samples.
\begin{figure}[t]
\centering
\includegraphics[width=0.4\textwidth]{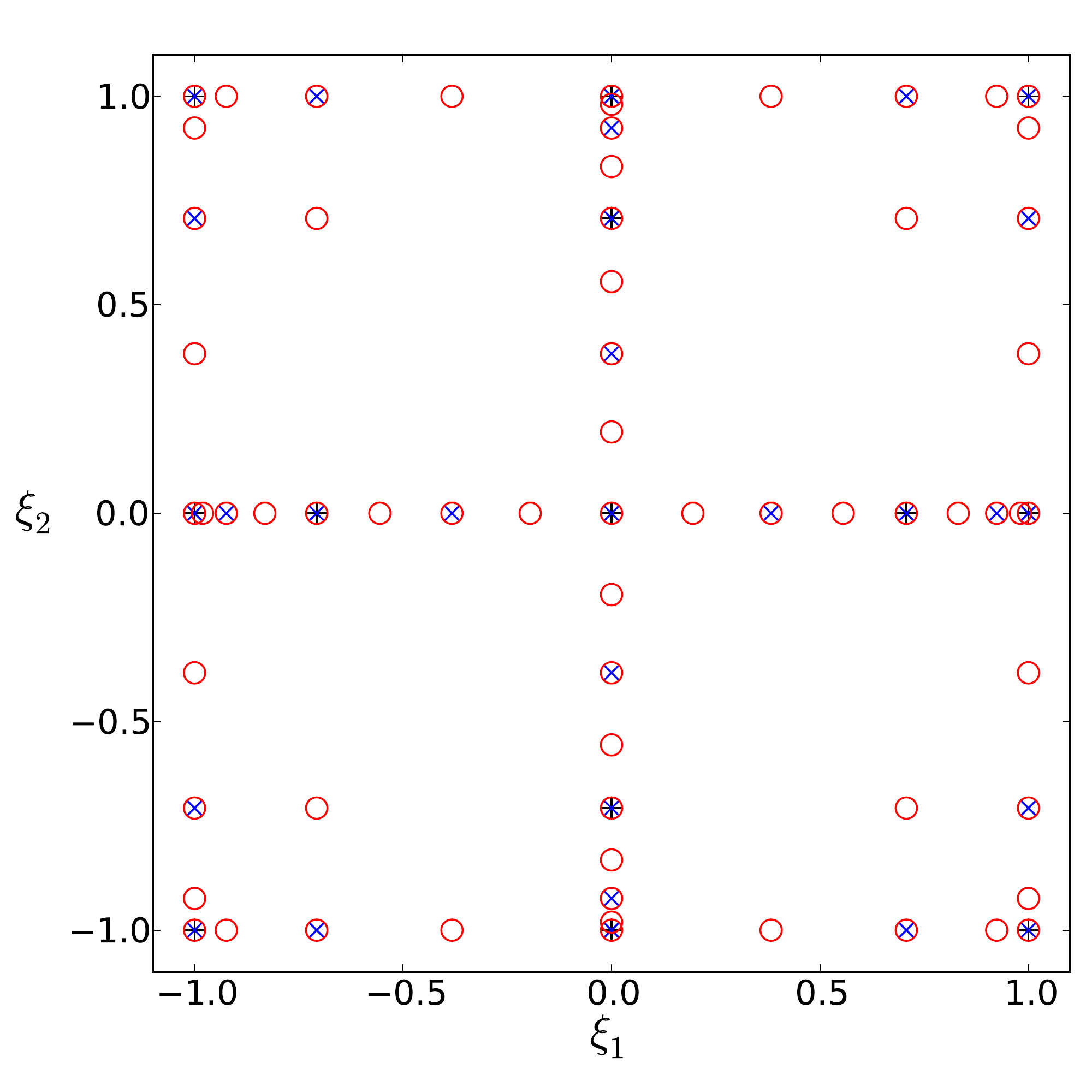}
\caption{\label{fig:sgr} Placement of deterministic samples via a
  sparse grid approach. }
\end{figure}
Figure~\ref{fig:sgr} shows the locations of deterministic samples we use, with a
sparse grid employing Clenshaw-Curtis quadraure. Several nested levels are shown
in the figure, with ``+'', ``x'', and ``o'' markers to illustrate the fact that
model evaluations can be re-used if higher-order approximations are necessary.
The number of requisite samples using sparse-quadrature evaluation of the
projection integrals, for a given requisite surrogate accuracy, is much smaller
than the corresponding number of MC samples, as we now illustrate.

\section{Numerical Results}
\label{sec:results}

\begin{figure}[h]
\centering 
\includegraphics[clip,width=0.35\textwidth]{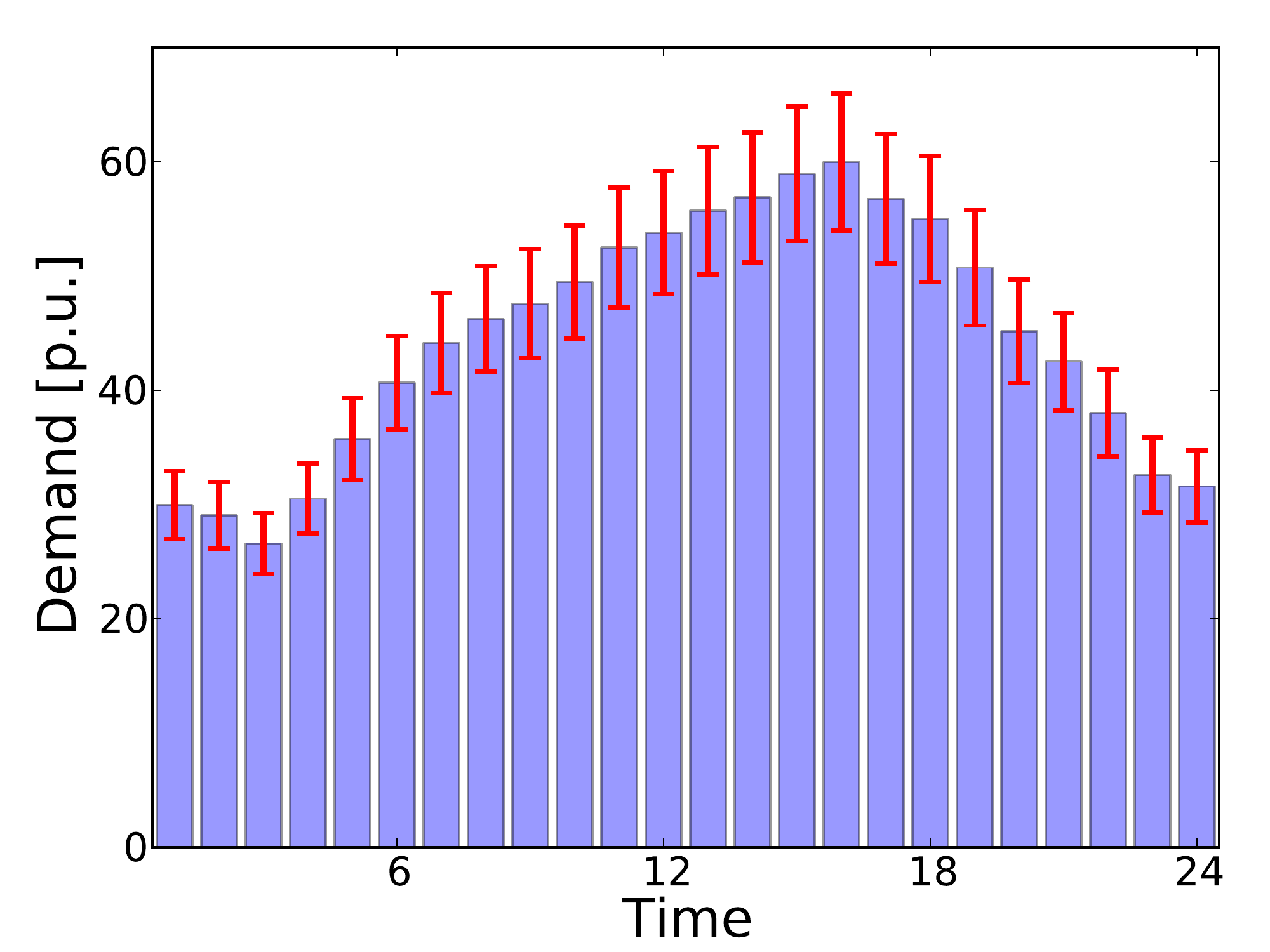} \\
\caption{\label{fig:dmd} Uncertain demand values for the IEEE 118-bus
  example. The error range is 20\% around the nomical values.}
\end{figure}

We now present numerical results comparing the calculation of the 
expected cost of ED via our proposed surrogate model with the
classical approach via Monte Carlo scenario sampling. We consider
two cases: a 9-bus example~\cite{Chow:1982} and the IEEE 118-bus
test system~\cite{IEEEdata}. We vary the number of time periods $|T|$
from 6 to 24. Following the model and analysis in \cite{carrionarroyo06}, 
we relax transmission constraints and focus on generating unit characteristics.
In Figure~\ref{fig:dmd}, we show a typical load series for the IEEE 118 
bus test system. The red bars depict a $20\%$ uncertainty range around 
the nominal values.

We first proceed to test the accuracy of the PCE surrogate for
$Q(\bm{x},\bm{\xi}(\bm{D}))$ with respect to full model evaluation. For this exercise we 
select the
9-bus model since the full model evaluation is expensive, hence we can
fully explore the high-dimensional demand space. Table~\ref{tab:l2err} shows the relative
$L_2$ error computed with the discrepancies observed at quadrature
points between truncated PCEs, in Eq.~(\ref{eq:qpce}), and the cost
values via direct evaluations. Several sparse quadrature levels are
employed. The table shows results for PCE orders from $1$ through
$4$. A three-level sparse quadrature, denoted L3 in the table, is sufficient
to construct a second-order PCE model with a negligible error compared
to the full model. Moreover, depending on the purpose of the PCE
model, a first-order approximation can also be
sufficient. Cross-validation tests using ensembles with $10^6$ random samples,
results not shown, confirm the high accuracy of the PCE model.
\begin{table}[t]
\centering
\begin{tabular}{c|c|c|c|c|} 
\hline
\multicolumn{1}{ |c| }{\multirow{2}{*}{Order} } & \multicolumn{4}{ c|
}{Sparse Quadrature} \\ \cline{2-5}
\multicolumn{1}{ |c| }{} & L2, 85p &  L3, 389p &  L4, 1457p &  L5, 4865p \\ \hline
\multicolumn{1}{ |c| }{1} & 1.62e-05 & 2.90e-05 & 2.15e-05 &  2.18e-05\\
\multicolumn{1}{ |c| }{2} & -              & 7.48e-07 & 2.17e-07 &  7.83e-08\\
\multicolumn{1}{ |c| }{3} & -              & -              & 1.92e-07 &  5.36e-08\\
\multicolumn{1}{ |c| }{4} & -              & -              & -              &  2.10e-08\\ \hline
\end{tabular}
\caption{Relative $L_2$ error at training points for several PCE
  surrogates and sparse quadrature levels. Power generation cost
  discretized using 10 segments.}
\label{tab:l2err}
\end{table}
Similar tests indicate that point-wise discrepancies between the PCE surrogate
and the full model are less than $0.5\%$ throughout the computational domain for a
2nd order PCE.

\begin{figure}[h]
\centering 
\includegraphics[trim=20mm 20mm 20mm
20mm,clip,width=0.37\textwidth]{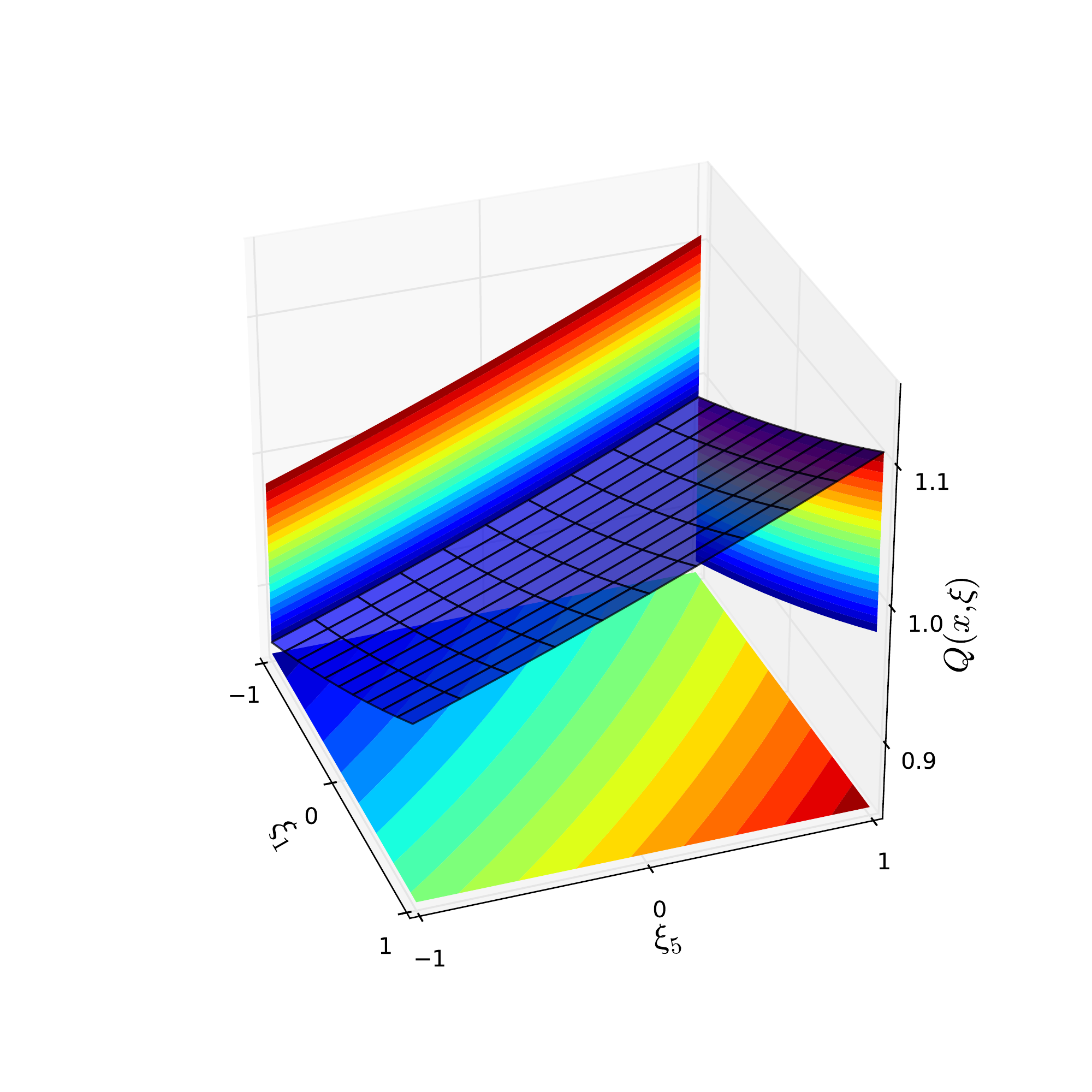} \\
\includegraphics[trim=20mm 20mm 20mm
20mm,clip,width=0.37\textwidth]{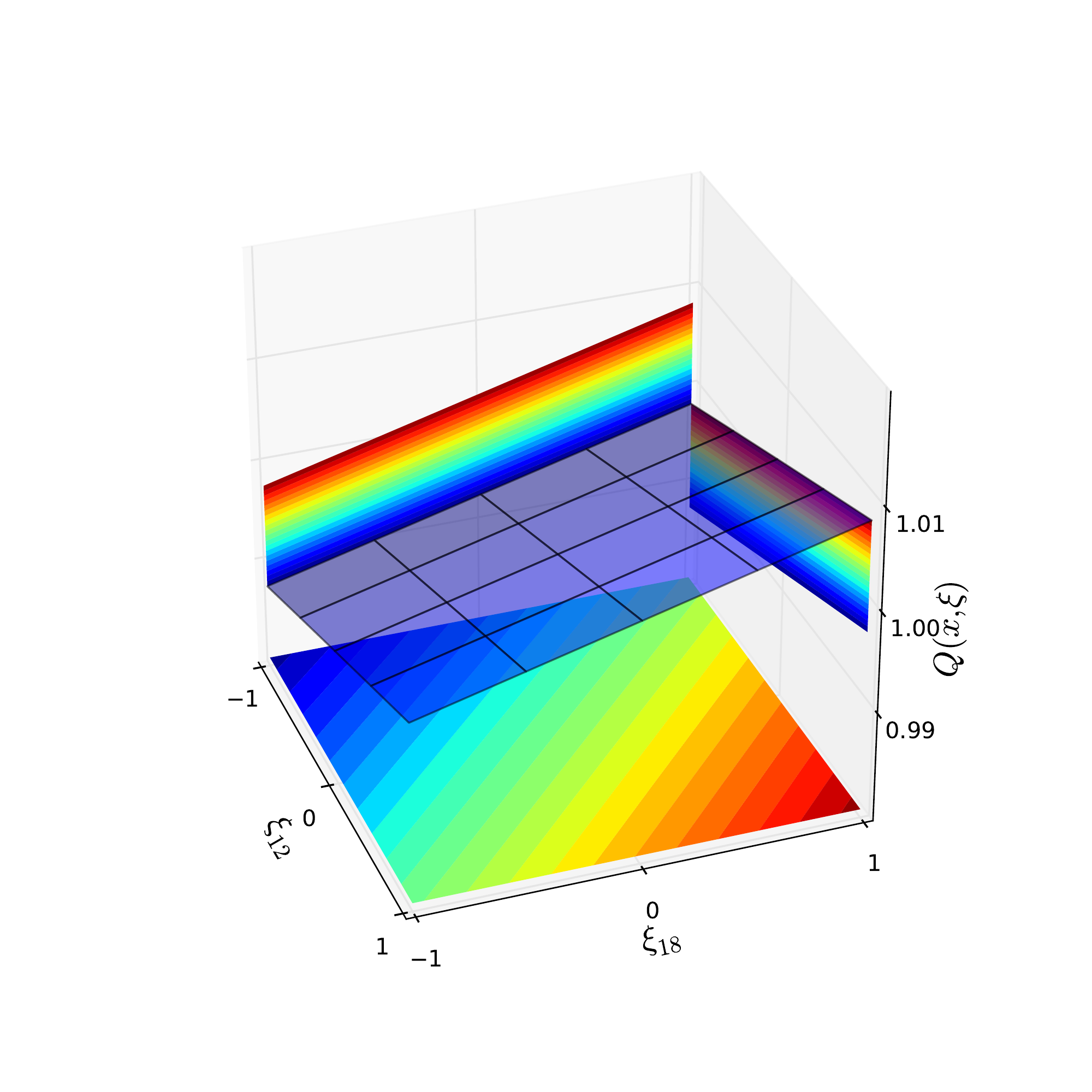} 
\caption{\label{fig:pcslice} Top frame: Slice through a 6-dimensional
  2nd order PCE model for the 9-bus system. Bottom frame: Slice
  through a 24-dimensional PCE model for the 118-bus system. The
  transparent surface shows the 3D dependence of $Q$ on the
  corresponding loads. Filled contours are projected on the side
  planes to provide a qualitative view of the dependence on each
  load.}
\end{figure}

Figure~\ref{fig:pcslice} illustrates the dependence of cost on the
load in specific time periods. This figure shows 2-dimensional slices
through the load spaces for the 9-bus (top frame) and 118-bus (bottom
frame) models. The cost $Q$ in these graphics is normalized by the
expected value for each case, and $\xi_t$ is computed from $D^t$ using Eq.~(\ref{eq:D2xi}).

\begin{figure}[t]
\centering
\includegraphics[width=0.35\textwidth]{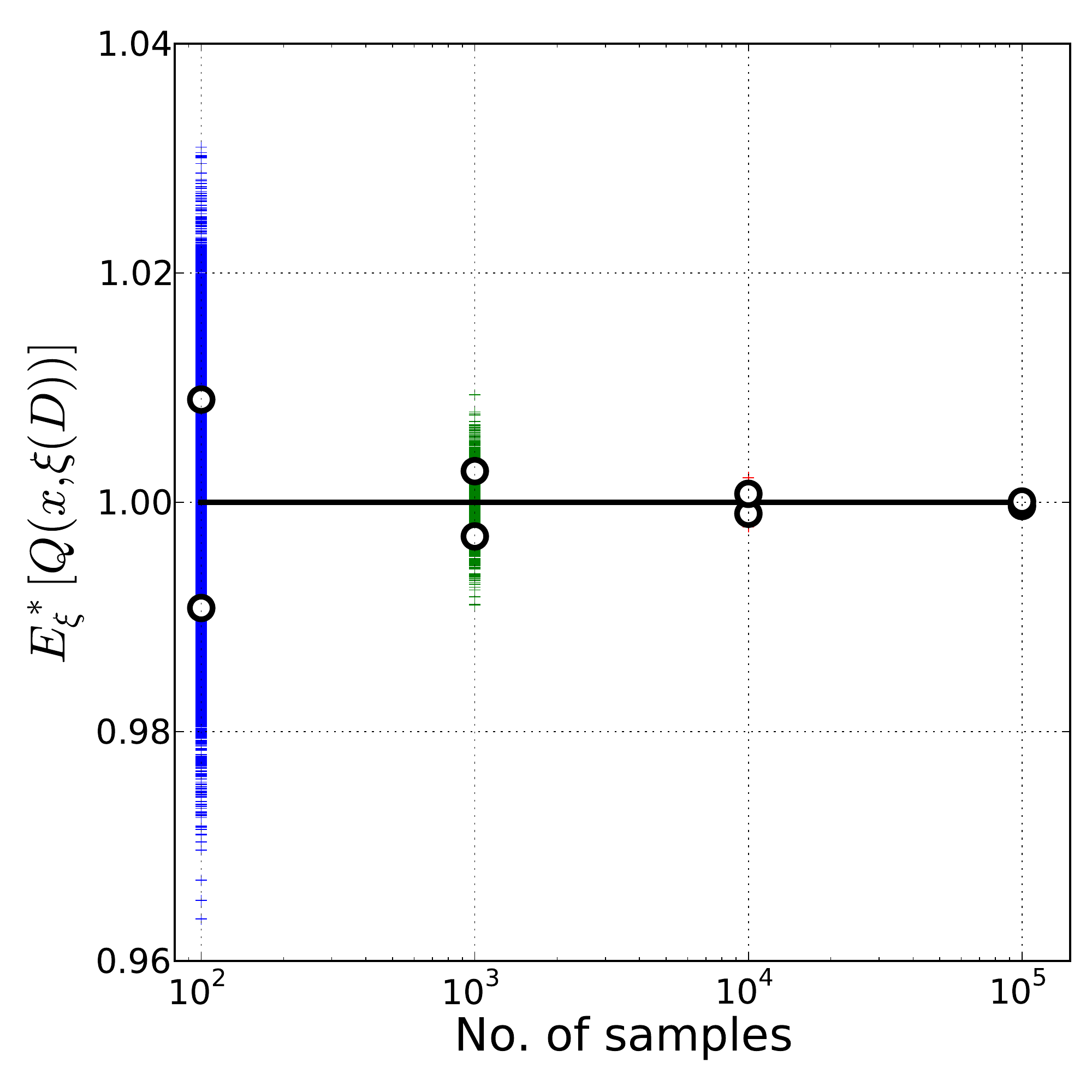}
\caption{\label{fig:pcst} Convergence of $\overline Q(\bm{x})=\mathbb E_{\bm{\xi}}
  Q(\bm{x}, \bm{\xi}(\bm{D}))$ computed via MC vs PCE.
  Scatter plots show normalized $\mathbb E_{\bm{\xi}}
  Q(\bm{x}, \bm{\xi}(\bm{D}))$ computed an increasing number of MC
  samples. The filled circles show $\pm\sigma$ for each ensemble, and
  the horizontal line shows the values computed via PCE. The
  expectation values are normalized by the ``true'' value obtained by
  MC using $10^6$ samples.} 
\end{figure}

Our research goal in this paper is to demonstrate the efficiency of using
PCE to compute (approximate)
the expected cost over all possible load realizations. 
Figure~\ref{fig:pcst} shows the convergence of the standard deviation of
the expected cost for the 9-bus model with 6 load periods, evaluated
using a number of scenarios between $10^2$ and $10^5$. These results
indicate that about $10^4$ samples are necessary to reduce the
relative error of the expected cost to below $10^{-3}$. 

\begin{figure}[t]
\centering
\includegraphics[width=0.35\textwidth]{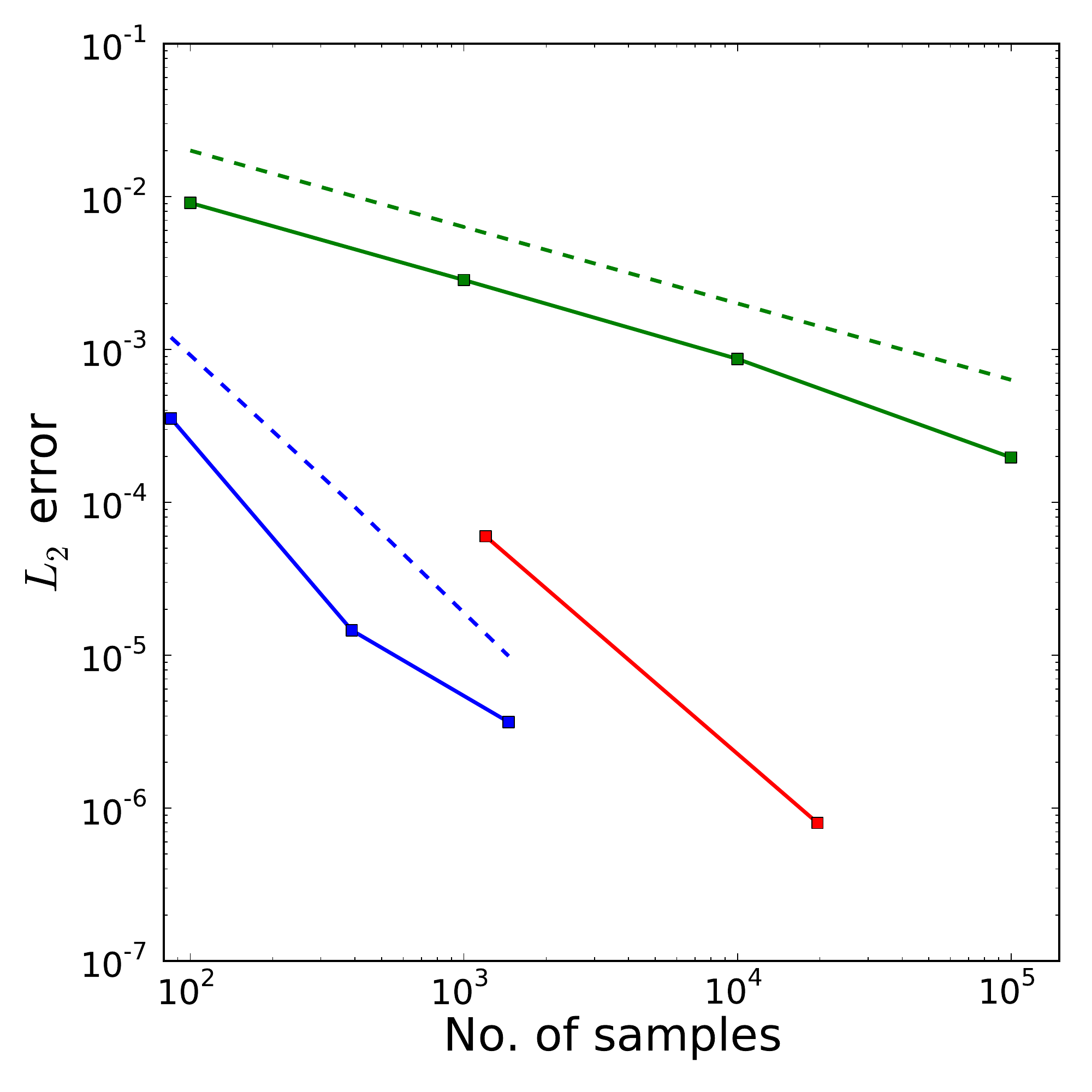}
\caption{\label{fig:conv} Convergence of $\overline Q(\bm{x})=\mathbb E_{\bm{\xi}}
  Q(\bm{x}, \bm{\xi}(\bm{D}))$ computed via MC (solid green line)
  vs PCE. The results for a 6-dimensional PCE corresponding to
  the 9-bus case with 6 load periods is shown with solid blue line, while the results for
  a 118-bus case with 24 load perionds and the equivalent 24-dimensional PCE is
  shown with solid red line. The dashed lines show theoretical
  convergence rates of $1/2$ (green) and $2$ (blue).}
\end{figure}

Figure~\ref{fig:conv} shows sample convergence rates for MC
compared to the PCE approach. For MC, a relative error of
$10^{-3}$ requires approximately $10^4$ model evaluations. The number of model
evaluations is independent of the number of loads periods, and is
nearly the same for the 9-bus and 118-bus systems. For the
surrogate model approach, the number of model evaluations represents
the cost of building the PCE. This cost is higly dependent on the
dimensionality of the surrogate. For example, the number of model
evaluations for a 24-dimensional PCE is one to two orders of magnitude
larger compared to a 6-dimensional PCE. Once the PCE is constructed,
subsequent evaluations of $\overline Q(\bm{x})$
incur negligible cost.
Hence the efficiency of the surrogate model approach over the
routine MC approach is proportional to the number of times  $\overline
Q$ needs to be evaluated during a simulation. For the
118-bus model, $10^3$ evaluations of Eq.~(\ref{buc}) results in a
computational time about $10^4$ times smaller for the PCE approach
compared to the traditional MC approach.

\section{Conclusion}
\label{sec:conclusion}

In this paper, we present an approach to reduce the computational cost
associated with stochastic unit commitment and economic dispatch, by
reducing the number of required forecast samples.
This approach is based on surrogate
models for the generation cost that cover the uncertainty of forecast
load. The surrogate models are constructed using Polynomial Chaos
Expansions. The construction of the terms in the surrogate models is
based on the projection of the model on increasingly higher basis
modes. Consequently, the global error in an $L_2$ sense between the surrogate
model and the actual simulations is easily controlled.

We present computational results using 9-bus and 118-bus test cases. For
both of these cases, quadratic surrogate models for the generation cost
showed global $L_2$ errors less than $10^{-4}$ while pointwise errors
were less than $1\%$ throughout the uncertain demand space. The
construction of Polynomial Chaos surrogate models typically requires
a much smaller number of samples, typically one to two orders of
magnitude for the examples considered in this paper,
compared to Monte Carlo evaluation of the expected generation cost for
a given requisite accuracy.
Subsequent evaluation of the
generation cost statistics via surrogate models incurs negligible
additional cost, thereby potentially reducing the computational
expense of the forecast ensembles in a stochastic unit commitment and
economic dispatch by several orders of magnitude.

\ifCLASSOPTIONcompsoc
  \section*{Acknowledgments}
\else
  \section*{Acknowledgment}
\fi
This work was funded by the Laboratory Directed Research \&
Development (LDRD) program at Sandia National Laboratories. Sandia
National Laboratories is a multiprogram laboratory operated by Sandia
Corporation, a wholly owned subsidiary of Lockheed Martin Corporation,
for the United States Department of Energy's National Nuclear Security
Administration under contract DE-AC04-94AL85000.

\ifCLASSOPTIONcaptionsoff
  \newpage
\fi


\end{document}

%% file: def.tex
\def\1{\ensuremath{_1}}
\def\2{\ensuremath{_2}}
\def\3{\ensuremath{_3}}
\def\4{\ensuremath{_4}}
\def\5{\ensuremath{_5}}
\def\6{\ensuremath{_6}}
\def\7{\ensuremath{_7}}
\def\8{\ensuremath{_8}}

\newcommand{\gae}{\lower 2pt \hbox{$\, \buildrel {\scriptstyle >}\over {\scriptstyle \sim}\,$}}
\newcommand{\lae}{\lower 2pt \hbox{$\, \buildrel {\scriptstyle <}\over {\scriptstyle \sim}\,$}}

\newcommand{\be}{\begin{equation}}
\newcommand{\ee}{\end{equation}}
\newcommand{\benn}{\begin{equation*}}		
\newcommand{\eenn}{\end{equation*}}
\newcommand{\bea}{\begin{eqnarray}}
\newcommand{\eea}{\end{eqnarray}}
\newcommand{\bean}{\begin{eqnarray*}}
\newcommand{\eean}{\end{eqnarray*}}

\newcommand\RR{{\rm I\kern-.16em R}}
\newcommand{\invisible}[1]{ }

\newcommand{\bi}{\begin{itemize}}
\newcommand{\ei}{\end{itemize}}

\newcommand{\etavec}{{\boldsymbol \eta}}

\newcommand{\Sfr}{\mathfrak{S}}
\newcommand{\xivec}{{\boldsymbol{\xi}}}
\newcommand{\xvec}{{\boldsymbol{x}}}